\documentstyle[tighten,preprint,aps,prb]{revtex}  
\begin{document}
\draft
\title{Interplay of charge and orbital ordering in manganese perovskites}
\author{K. H. Ahn and A. J. Millis}
\address{Department of Physics and Astronomy, The Johns Hopkins University
\\
Baltimore, Maryland 21218}
\maketitle
\begin{abstract}
A model of localized classical electrons coupled to
lattice degrees of freedom and, via the Coulomb 
interaction, to each other, has been studied 
to gain insight into the charge and orbital ordering observed in 
lightly doped manganese perovskites.
Expressions are obtained for the minimum energy and ionic displacements 
caused by given hole and electron orbital configurations.
The expressions are 
analyzed for several hole configurations,
including that 
experimentally observed by Yamada {\it et al.}
in ${\rm La_{7/8}Sr_{1/8}MnO_3}$.
We find that, 
although the preferred charge and orbital ordering 
depend sensitively on parameters, 
there are ranges of the parameters
in which the experimentally observed hole configuration
has the lowest energy. 
For these parameter values we also find that 
the energy differences between different hole configurations are 
on the order of the observed charge ordering transition temperature. 
The effects of additional strains are also studied.
Some results for ${\rm La_{1/2}Ca_{1/2}MnO_3}$ are presented, 
although our model may not adequately describe this material 
because the high temperature phase is metallic.
\end{abstract}
\narrowtext

Over the last few years
much attention has been focused on manganese perovskite-based oxides, 
most notably the pseudocubic materials 
$Re_{1-x}Ak_x{\rm MnO_3}$. (Here $Re$ is a rare earth element such as La,
and $Ak$ is a divalent alkali metal element such as Ca or Sr.)
The initial motivation came  from the observation that 
for some range of $x$, and temperature, $T$, 
resistance can be reduced by a factor of up to $10^7$ 
in the presence of a magnetic field.\cite{ahn} 
Two other interesting physical phenomena 
occurring in this class of materials are
charge ordering and orbital ordering.\cite{chen}
In this paper, we study the connection between the two.

The important electrons in $Re_{1-x}Ak_x{\rm MnO_3}$ are the Mn $e_g$ 
electrons; their concentration is $1-x$.
For many choices of $Re$, $Ak$, and $x$, 
especially at commensurate $x$ values, 
the $e_g$ charge distribution is not uniform
and it indeed appears that
a fraction $x$ of Mn ions have no $e_g$ electron 
while $1-x$ have a localized  $e_g$ electron.
A periodic pattern of filled and empty sites is 
said to exhibit charge ordering. 
There are two $e_g$ orbitals per Mn ion.   
A localized Mn $e_g$ electron 
will be in one linear combination of these; 
a periodic pattern of orbital occupancy is said to exhibit 
orbital ordering.
Recently, Murakami {\it et al.} \cite{murakami} 
observed the charge ordering transition
accompanying simultaneous orbital ordering 
in ${\rm La_{1/2}Sr_{3/2}MnO_4}$
at 217 K
(well above the magnetic phase transition temperature 110 K).
It indicates that the interplay of the charge and orbital ordering
to minimize the lattice energy
could be the origin of the charge ordering.
In this paper we present an expression for the coupling
between charge and orbital ordering, 
with different charge ordering patterns
favoring different orbital orderings. 
We also argue that the orbital ordering energy differences       
determine the observed charge ordering in lightly doped manganites.
Localized charges induce local lattice distortions, 
which must be accommodated into the global crystal structure;
the energy cost of this accommodation is 
different for different charge ordering patterns.

To model the charge and orbital ordering, 
we assume that the electrons are localized classical objects,
so that each Mn site is occupied by zero or one $e_g$ electron,
and each $e_g$ electron is in a definite 
orbital state.
This assumption seems reasonable in the lightly doped materials such as 
${\rm La_{7/8}Sr_{1/8}MnO_3}$,
which are strongly insulating at all temperatures,\cite{yamada}
but may not be reasonable for
the ${\rm La_{1/2}Ca_{1/2}MnO_3}$ composition,\cite{chen}
where the charge ordered state emerges at a low temperature 
from a metallic state.
We proceed by calculating the energies of different charge ordering
patterns, emphasizing the 1/8 doping case.
It is practically impossible to consider all
possible charge ordering configurations.
Therefore, we consider the three configurations
shown in Fig.\ \ref{fig1},
which are the only ones consistent with the  following
basic features of the hole-lattice implied by the experimental results
by Yamada {\it et al.}\cite{yamada} :
invariance under translation by two lattice constants
in the $x$ or $y$ direction, four in the $z$ direction, and an alternating 
pattern of occupied and empty planes along $z$ direction.
The configuration in
Fig.\ \ref{fig1}(b) is the one proposed 
by Yamada {\it et al.}\cite{yamada} 
to explain their experimental results for  ${\rm La_{7/8}Sr_{1/8}MnO_3}$. 
For localized electrons there are three energy terms :
the coupling to the lattice,
which will be discussed at length below, 
the Coulomb interaction, 
and the magnetic interaction. 

First, we argue that the Coulomb energy cannot 
explain the observed ordering pattern or transition 
temperature.
We take as reference the state with one $e_g$ electron per Mn 
and 
denote by 
$\delta q_i$ the charge of a hole on a Mn site.  
From the classical Coulomb energy 
\begin{equation}
U_{\rm Coulomb}=\frac{1}{2\epsilon_0}
\sum_{i \neq j}
\frac{\delta q_i \delta q_j}{r_{ij}},
\end{equation}
one finds that 
the difference in energy between
the  configurations in Fig.\ \ref{fig1} is
\begin{equation}
\Delta{\cal U}_{\rm Coulomb,\: per\: hole} =\frac{1}{2 \epsilon_0}
\sum_{i\neq o}
\frac{\Delta(\delta q_i)}{r_{io}},
\end{equation}
where $o$ is a site containing a hole and 
$\Delta(\delta q)$ is the difference in charge 
between the two configurations.
We estimated the above infinite sum by repeated numerical calculations for
larger and larger volumes of the unit cells around the origin. 
We find that Fig.\ \ref{fig1}(c) has the lowest  energy; 
12 meV/$\epsilon_o$ lower
than Fig.\ \ref{fig1}(b), and 27 meV/$\epsilon_o$  
lower than Fig.\ \ref{fig1}(a).

To estimate the magnitude of the Coulomb energy differences, 
we need an estimate for the dielectric constant $\epsilon_0$,
which we obtain from 
the measured reflectivity 
for 
${\rm La_{0.9}Sr_{0.1}MnO_3}$,\cite{okimoto1}
and  
the Lyddane-Sachs-Teller relation\cite{aschcroft}
$\omega_L^2=\omega_T^2 \epsilon_0 / \epsilon_{\infty}$.
At frequencies greater than the greatest phonon frequency 
the reflectivity is close to 0.1,
implying 
$\epsilon_{\infty} \approx 3.4$;
the reflectivity is near unity between 
$\omega_T=0.020$ eV, and
$\omega_L=0.024$ eV, 
implying $\epsilon_0 \approx 5.0$.
Because both ${\rm La_{7/8}Sr_{1/8}MnO_3}$ and 
${\rm La_{0.9}Sr_{0.1}MnO_3}$
are insulating and have similar compositions,
their static dielectric constants are expected to be similar. 
Using $\epsilon_0\approx 5.0$,
the energy difference between different configurations of holes 
is only around 2.4 meV, or 30 K per hole, which
is small compared to the observed charge ordering temperature of 
150 K $-$ 200 K of these materials.
The inconsistency with the experimentally observed hole configuration and 
the smallness of the energy difference
scale indicate that the electrostatic energy 
is not the main origin of charge ordering 
for this material.

Even though the magnetic and charge ordering transitions 
show a correlation in 
 ${\rm La_{7/8}Sr_{1/8}MnO_3}$,\cite{kawano}
we  do not think that the magnetic contribution to charge and orbital
ordering is as important as the 
lattice contribution for three reasons. 
First, in undoped ${\rm LaMnO_3}$, 
the orbital ordering and the structural phase transition 
occur at around 
800 K and the magnetic ordering at around 140 K,\cite{wollan,kanamori} 
suggesting that the magnetic effects are relatively weak.
Second, in ${\rm La_{7/8}Sr_{1/8}MnO_3}$ the Mn spins are
ferromagnetically 
ordered with moment close to the full Mn moment 
at temperatures greater than
the charge ordering temperature,\cite{kawano}
 and ferromagnetic order does not favor 
one charge configuration over another.
Third, although in ${\rm La_{7/8}Sr_{1/8}MnO_3}$ antiferromagnetic 
order appears at the charge ordering transition,
the antiferromagnetic 
moment is very small (less than 0.1 of the full Mn moment),\cite{kawano} 
so the energy associated 
with this ordering must be much less than 140 K/site associated with 
magnetic ordering in ${\rm LaMnO_3}$.
Therefore, we think that the canted antiferromagnetism occuring upon
charge ordering in ${\rm La_{7/8}Sr_{1/8}MnO_3}$ (Ref.\ 7)
is not the cause but the effect of the charge and orbital ordering.
We now turn our attention to the lattice energy.

A classical model for the lattice distortions of the insulating perovskite
manganites has been derived in Ref.\ 10, 
and shown to be consistent with experimental results on ${\rm LaMnO_3}$.
This  model is adopted here
with an
additional term,
an energy cost for shear strain.
We now briefly outline the model, 
which is explained in more detail 
in Ref.\ 10 and the Appendix.
The ionic displacements included are 
the vector displacement $\vec{\delta}_i$ of the Mn ion
on site $i$, 
and the $\hat{a}$ directional scalar displacement 
$u_i^a$ ($a=x$, $y$, and $z$)
of the O ion 
which sits between the Mn ion on site $i$ and 
the Mn ion on site $i+\hat{a}$.
For convenience, $\vec{\delta}_i$ and $u_i^a$ 
are defined to be dimensionless in the following way:
the lattice constant of the ideal cubic perovskite is $b$,
the Mn ion position in the ideal cubic perovskite is $\vec{R}_i$,
the actual Mn ion position is $\vec{R}_i+b\vec{\delta}_i$,
and the actual O ion position is 
$\vec{R}_i+(b/2+bu_i^a)\hat{a}$.    
The lattice energy is taken to be harmonic and depends only 
on the nearest neighbor   
Mn-O distance and 
the first and second nearest neighbor 
Mn-Mn distances.
The spring constants corresponding to these displacements are
$K_1$, $K_2$, and $K_3$ as shown in Fig.\ \ref{fig2}.
Because $K_1$ and $K_2$ involve bond stretching,
while $K_3$ involves bond bending,
$K_1\geq K_2 \gg K_3$ is expected.
Thus, 
$E_{\rm lattice}=E_{\text{Mn-O}}
+E_{\text{Mn-Mn,first}}+E_{\text{Mn-Mn,second}}$,
where
\begin{eqnarray}
 E_{\text{Mn-O}} &=& \frac{1}{2} K_1 
\sum_{i,a}(\delta^{a}_{i}-u^{a}_{i})^{2}
+(\delta^{a}_{i} - u^{a}_{i-a})^{2}, \\
 E_{\text{Mn-Mn,first}} &=&  \frac{1}{2} K_2 
  \sum_{i,a}(\delta^{a}_{i}-\delta^{a}_{i-a})^{2}, \\
 E_{\text{Mn-Mn,second}} &=& \frac{1}{2} K_3  
\sum_{
i,(a,b)  
}
\left[ \left(\frac{\delta^a_{i+a+b}+\delta^b_{i+a+b}}{\sqrt{2}}\right)-
  \left(\frac{\delta^a_{i}+\delta^b_{i}}{\sqrt{2}}\right) \right]^{2} 
  \nonumber \\
& &+  
\left[ \left(\frac{\delta^a_{i+a-b}-\delta^b_{i+a-b}}{\sqrt{2}}\right)-
  \left(\frac{\delta^a_{i}-\delta^b_{i}}{\sqrt{2}}\right)\right]^{2}. 
\end{eqnarray}
In the above equations $a$ denotes 
$x$, $y$, and $z$, and $(a,b)$ represents
$(x,y)$, $(y,z)$, and $(z,x)$.
$E_{\text{Mn-Mn,second}}$  
was
not considered in Ref.\ 10.
The shear modulus produced by this term 
is important, because without it,
a Mn ion on site $i+\hat{x}$ can 
have arbitrary large $y$ directional displacement 
relative to the Mn ion on site $i$ at no cost of energy.
For this reason,
the model with $K_3=0$  has singularities,
whose proper treatment requires $K_3\neq 0$ in our model.
However, still we expect $K_3$ will be much smaller than $K_1$ or $K_2$. 
Therefore, in order to simplify the calculation,
 the $K_3/K_1\rightarrow0$ limit has been taken after the expression
of minimized energy and equilibrium ionic displacements
have been obtained.

Second, we consider the electronic degree of freedom.
We parameterize the electron density by the variable
$h_i$.
If an electron is present on site $i$, 
$h_i=0$;
if no electron is present,
$h_i=1$.
If there is an electron on site $i$, 
the electron orbital state, 
which is a linear combination of the two $e_g$ orbitals,
is parameterized by an angle $\theta_i$ as follows. 
\begin{equation}
|\psi_{i}(\theta_i)>=
\cos{\theta_{i}}|d_{3z^{2}-r^{2}}>+\sin{\theta_{i}}|d_{x^{2}-y^{2}}>
\end{equation}
with $0 \leq \theta_{i} < \pi$.
The electron orbital state couples 
to the distortion of the surrounding
oxygen octahedra through the Jahn-Teller distortion.
The coupling is given by   
\begin{eqnarray}
E_{\rm JT} &=& - \lambda \sum_{i} (1-h_{i})[\cos{2\theta_{i}}\{v^{z}_{i}-
  \frac{1}{2}(v^{x}_{i}+v^{y}_{i})\}+
\sin{2\theta_{i}}\frac{\sqrt{3}}{2}(v^{x}_{i}-v^{y}_{i})] \nonumber \\
 &=& -\lambda\sum_{i,a}(1-h_{i})v^{a}_{i} \cos{2(\theta_{i}+\psi_{a})}, 
\end{eqnarray}
where
\begin{eqnarray}
v^{a}_{i}&=&u^{a}_{i}-u^{a}_{i-a},  \\
\psi_{x}=-\pi/3,\: \psi_{y}&=&\pi/3,\: \psi_{z}=0.
\end{eqnarray}
If a hole is present on site $i$, 
it attracts the surrounding oxygens equally,
giving rise to a breathing distortion energy given by 
\begin{equation}
E_{\rm hole}=\beta\lambda\sum_{i}h_{i}(v^{x}_{i}+v^{y}_{i}+v^{z}_{i}).
\end{equation}
The parameter $\beta$ represents the strength
of the breathing distortion 
relative to
the Jahn-Teller distortion.
Finally, following Kanamori,\cite{kanamori} 
we include a phenomenological cubic anharmonicity term given by
\begin{equation}
E_{\rm anharm} = - A \sum_{i} (1-h_{i}) \cos{6\theta_i}. \label{eq:anharm}
\end{equation}
The sign has been chosen so that the electron orbital states of
$|3x^2-r^2>$, $|3y^2-r^2>$, or $|3z^2-r^2>$, with $\hat{x}$, $\hat{y}$, 
and $\hat{z}$ pointing
toward nearest oxygen ions 
are favored when $A$ is positive.
The total energy, which is the sum of all the above energy terms, 
is given by 
\begin{equation}
E_{\rm tot} = E_{\text{Mn-O}}+E_{\text{Mn-Mn,first}}+
E_{\text{Mn-Mn,second}}+E_{\rm JT}+E_{\rm hole}+E_{\rm anharm}.
 \label{eq:TOT}
\end{equation}

We minimized $E_{\rm tot}$ about $\delta_i^a$'s and $u_i^a$'s for
fixed hole and orbital configurations.
These are conveniently expressed in terms of the variables
$\delta_{\vec{k}}^a$, $u_{\vec{k}}^a$,  
$h_{\vec{k}}$, and $c_{\vec{k}}^a$ defined in the following way.
\begin{eqnarray}
\delta_{i}^{a} &=& \sum_{\vec{k}} 
e^{-i \vec{k} \cdot \vec{R}_{i}}\delta_{\vec{k}}^{a}, \label{eq:ee}\\
u_{i}^{a} &=& \sum_{\vec{k}} e^{-i \vec{k} 
\cdot \vec{R}_{i}} u_{\vec{k}}^{a}, \label{eq:uu}\\
h_{i} &=& \sum_{\vec{k}} e^{-i \vec{k} \cdot \vec{R}_{i}} 
h_{\vec{k}}, 
\label{eq:hole} \\
(1-h_{i})\cos{2(\theta_{i}+\psi_{a})} &=& \sum_{\vec{k}} e^{-i \vec{k} 
\cdot \vec{R}_{i}} c_{\vec{k}}^{a}.  \label{eq:orbital} 
\end{eqnarray}
The details are shown in the Appendix.
The minimized energy per Mn ion may be written as
\begin{equation}
\frac{E_{\rm tot}}{N}={\cal E}_{\vec{k}=0}+
\sum_{\vec{k}\neq 0, a}{\cal E}^a_{\vec{k}}+
      \frac{E_{\rm anharm}}{N}, \label{eq:tot}
\end{equation}
where
\begin{eqnarray}
{\cal E}^a_{\vec{k}}&=&
\left\{ \begin{array}{ll}
           -\frac{\lambda^2}{(K_1+2K_2)K_1} [ K_1+K_2(1-\cos{k_a}) ]
           (\beta h_{\vec{k}}-c^{a}_{\vec{k}})(\beta h_{-\vec{k}} 
- c^a_{-\vec{k}}), &
            \mbox{if $k_a\neq 0$ } \\
           0, & \mbox{if $k_a= 0$ }
       \end{array} 
\right. \label{eq:nonzerok} \\
{\cal E}_{\vec{k}=0}&=&-\frac{\lambda^2}{K_1+2K_2}[3(\beta h_0)^2
+\sum_{a}(c^a_0)^2]. 
\label{eq:zerok}
\end{eqnarray}
The long wave length strain $e^{ab}$, and the 
$\vec{k}(\neq 0)$ components of the ionic displacements
are given as 
\begin{eqnarray}
e^{ab}&=&-\frac{2\lambda}{K_1+2K_2}(\beta h_0-c_0^a) \delta_{ab}, \\
u^a_{\vec{k}\neq 0}&=&
   \left\{ \begin{array}{ll} 
             -\frac{\lambda [K_1+K_2(1-\cos{k_a})]}{(K_1+2K_2)K_1}
\frac{1-e^{-ik_a}}{1-\cos{k_a}}
             (\beta h_{\vec{k}} -c^a_{\vec{k}}), &
             \mbox{if $k_a\neq0$ } \\
             0, & \mbox{if $k_a=0$ }
           \end{array}
   \right. \\
\delta^a_{\vec{k}\neq 0}&=&
   \left\{ \begin{array}{ll} 
              -i\frac{\lambda\sin{k_a}}{(K_1+2K_2)(1-\cos{k_a})}
(\beta h_{\vec{k}} -c^a_{\vec{k}}), &
              \mbox{if $k_a\neq 0$ } \\
              0, & \mbox{if $k_a=0$. }
           \end{array}
   \right. \label{eq:del}
\end{eqnarray} 

Because $h_i$'s and $(1-h_{i})\cos{2(\theta_{i}+\psi_{a})}$'s are bounded
by $\pm 1$, 
we cannot treat $h_{\vec{k}}$'s and $c^a_{\vec{k}}$'s 
as independent variables to minimize $E_{\rm tot}$.
Therefore, we minimize $E_{\rm tot}$ 
over the orbital variable $\theta_i$'s 
at fixed hole configurations;
the ground state is then the hole 
configuration of the lowest energy.

For ${\rm La_{7/8}Sr_{1/8}MnO_3}$,
we consider the three hole configurations shown in Fig.\ \ref{fig1}, 
each of which is a Bravais lattice,
with a unit cell containing one Mn site 
with a hole and seven Mn sites without holes.
The orbital configuration may be different 
in different unit cells of the lattice defined by the holes.
We consider the case where the orbital configuration is 
the same in each unit cell.
In addition to that,  we also consider 
all possible two sublattice symmetry breakings.
Therefore, we have seven (if no symmetry breaking) 
or fourteen  (if two-sublattice symmetry breaking) 
orbital variable $\theta_i$'s.
$E_{\rm tot}/N$ in Eq.\ (\ref{eq:tot}) for each configuration
is expressed in terms of those variables through
Eqs.\ (\ref{eq:anharm}), (\ref{eq:hole}), (\ref{eq:orbital}), 
(\ref{eq:nonzerok}), 
and (\ref{eq:zerok}), 
and is minimized about $\theta_i$'s.
For this minimization, we use the {\it FindMinimum} routine 
in {\it Mathematica}
in the following way: 
for each set of parameters,  and for each configuration,
we check the local minimal values by using 50 $-$ 200 
random starting values
of $\theta_i$'s.

According to Ref.\ 10,
$\lambda/K_1$ ranges over 0.04 $-$ 0.05, 
and $K_2/K_1$ is between 0 and 1.
$A/K_1$ ranges around 0.0002, 
and 
$K_1\approx 200$ eV.\cite{millis}
Recently, a local breathing distortion of $0.12$ $\AA$ has been
directly observed in ${\rm La_{0.75}Ca_{0.25}MnO_3}$.\cite{billinge}
The Jahn-Teller distortion 
is
estimated 
around 0.15 $\AA$
from the Mn-O distances of ${\rm LaMnO_3}$.\cite{ellemans}
This implies that the breathing distortion and 
the Jahn-Teller distortion in these materials
have similar order of magnitude, i.e.,
$\beta=O(1)$.
We varied $\beta$ in the range of 0 $-$ 10,
and $A/K_1$ in the range of 0 $-$ 0.00035, with $\lambda/K_1=0.045$,
$K_2/K_1=0.5$, and $K_1= 200$ eV. 
For each set of those parameters, 
the minimum energy per hole for each fixed hole configuration
in Fig.\ \ref{fig1} has been found.
By comparing them, we find the most favored hole configuration 
for each $\beta$ and $A/K_1$,
which is shown in Fig.\ \ref{fig3} as a plot in $\beta$-$A/K_1$ plane.

At large $\beta$ ($\gtrsim 7$), 
the configuration shown in Fig.\ \ref{fig1}(c) is the most favored, 
and that shown in Fig.\ \ref{fig1}(a) is the least favored.
This can be related to the fact that, 
in $y$-$z$ and $z$-$x$ directional planes, 
Fig.\ \ref{fig1}(c) has the most even distribution of holes, 
and Fig.\ \ref{fig1}(a) has the least even distribution.
For large $\beta$, the contraction of oxygen octahedra toward  
holes is strong, 
and an uneven distribution of holes generates larger strains and 
elevates minimum energies.  
Particularly, the square hole net squeezes the electron orbital 
at the center of the square 
along the direction perpendicular to the square plane.
In the cubic hole configuration of Fig.\ \ref{fig1}(a), 
the six squeezed electron orbitals
point  toward the cubic center, putting the electron orbital at the center
at high energy, which is consistent with our result 
that Fig.\ \ref{fig1}(a) has far higher minimum energies
than Figs.\ \ref{fig1}(b) and \ref{fig1}(c) in large $\beta$ limit. 

As $\beta$ is decreased into the range of 2 $-$ 5, 
the favored hole configuration becomes that of Fig.\ \ref{fig1}(b), 
which is the experimentally observed hole configuration.
We expect that 
the difference of the energy per hole between 
the ground state  hole configuration and the next lowest 
energy hole configuration 
corresponds approximately to 
the charge ordering 
temperature. 
The calculation results indicate that, 
when $\beta$ is in the range of 2.0 $-$ 2.5 
or around $5.0$ and $A/K_1=0.0002$, 
the charge ordering temperature
is around 100 $-$ 200 K, 
which is consistent with experimental results.
As $\beta$ is decreased further, the most favored hole 
configuration changes
further and the temperature difference scale decreases.

Figure \ref{fig3} also shows the tendency that 
the configuration of Fig.\ \ref{fig1}(c) becomes 
more favored as $A/K_1$ increases.
We think this occurs  
because the anharmonicity energy distorts the oxygen octahedra
tetragonally, which can be more easily accommodated 
by the tetragonal hole configuration of 
Fig.\ \ref{fig1}(c).  

In Table\ \ref{table1}, we have shown an example of the orbital states, 
ionic displacements, and uniform strains 
corresponding to the minimum energy configuration 
for Fig.\ \ref{fig1}(b) when 
$A/K_1=0.0002$, $\lambda/K_1=0.045$, $K_2/K_1=0.5$ and $\beta=2.5$.
$x$, $y$, and $z$ directions are shown in Fig.\ \ref{fig1}.
The nearest Mn-Mn distance is unit.
$(n_i^x,n_i^y,n_i^z)$ is defined  in such a way that 
$(n_i^x,n_i^y,n_i^z)+N_1(2,0,0)+2N_2(0,2,0)+2N_3(1,0,2)$'s 
and 
$(n_i^x,n_i^y,n_i^z)+N_1(2,0,0)+(2N_2+1)(0,2,0)+(2N_3+1)(1,0,2)$'s, 
where $N_1$, $N_2$, and $N_3$ are integers,
represent the coordinates of the sites indexed by $i$.
$\vec{k}=0$ parts of the ionic displacements have been subtracted 
to find the non-uniform parts of the displacements.  

The energy expressions 
in Eqs.\ (\ref{eq:nonzerok}) and (\ref{eq:zerok}) 
are adequate for bulk materials.
When the material is grown on a substrate as a thin film,
generally there is a strain generated by lattice mismatch 
between the film and the substrate materials.
To see the effect of this strain, 
we add a term proportional to $c_0^{a'}$ ($a'=x$, $y$, or $z$) 
to the energy,
which corresponds to an $a'$ directional strain.
Using a parameter $g$, we replace ${\cal E}_{\vec{k}=0}$ 
in Eq.\ (\ref{eq:zerok}) 
by the following expression:
\begin{equation}
{\cal E'}_{\vec{k}=0}=-\frac{\lambda^2}{K_1+2K_2}[3(\beta h_0)^2
+g c^{a'}_0+\sum_{a}(c^a_0)^2
]. 
\end{equation}
We repeated similar calculations 
to find the favored hole configurations 
for different values of 
the applied strain,
parameterized by $g$.
The applied strain breaks cubic symmetry.
Some of the hole configurations also break cubic symmetry.
For these cases the energy depends on the relative orientation 
of the strain and hole symmetry breakings.
We consider all possible orientations and 
find the lowest energy state.  
We have varied  $g$  between $-0.4$ and $0.4$, 
and $\beta$  between 0 and 7,
with 
 $A/K_1=0.0002$, $\lambda/K_1=0.045$, 
and $K_2/K_1=0.5$.
The results are shown  
as a phase diagram in $\beta$-$g$ plane
in Fig.\ \ref{fig4}.
It shows that Fig.\ \ref{fig1}(c) configuration 
is favored more as $|g|$ increases.
This feature can be understood in the following way.
For small $g$'s,  
the leading correction to the minimum energy 
for each hole configuration is
$ -\lambda^2  g \tilde{c^{a}_0} / (K_1+2K_2) $,
where  $\tilde{c^{a}_0}$ represents $c^a_0|_{g=0}$.
Therefore the configuration which has a larger $|\tilde{c^{a}_0}|$ 
will show greater
change in energy for a given $g$.
Since the hole configuration in Fig.\ \ref{fig1}(c) has 
tetragonal symmetry,
which is compatible with  the Jahn-Teller distortion,
it has the largest  $|\tilde{c^a_0}|$.
Therefore, as $|g|$ increases, Fig.\ \ref{fig1}(c) is more favored than 
Figs.\ \ref{fig1}(a) and \ref{fig1}(b).
Because the energy changes linearly with $g$, 
the phase boundaries are straight lines for small $g$,
and have cusps at $g=0$ ,
as shown in Fig.\ \ref{fig4}.
Typical variations of $e^{aa}$ corresponding to 
changing $|g|$ from 0 to 0.4 
are about 2 \%.  
The results indicate that the strain generated by substrates 
can change ordered hole configuration and  ordering temperature.
  
Our results indicate that the interaction 
of the electronic state and the lattice 
can be the origin of the charge ordering in this material,  
even though the details of the results are  
dependent on specific choice of
$K_1$, $K_2$, $\lambda$, and $A$.

A similar calculation has been done for $Re_{1/2}Ak_{1/2}{\rm MnO_3}$,
 hole concentration 1/2.
We choose the three hole configurations in Fig.\ \ref{fig5} 
to compare the minimum energies.
Each configuration has an alternating hole distribution 
in different set of directions :
$x$, $y$, and $z$ directions for Fig.\ \ref{fig5}(a), 
$x$ and $y$ directions for Fig.\ \ref{fig5}(b),
and $y$ direction for Fig.\ \ref{fig5}(c).
Figure\ \ref{fig5}(b) is the experimentally observed hole 
configuration.\cite{chen}
As we have done for $x=1/8$, we consider 
both the case
where hole and orbital state have the same unit cell,
and the case where the orbital state is composed 
of the two hole sublattices.
Calculations for $A/K_1=0.0002$, $\lambda/K_1=0.045$, and
$K_2/K_1=0.5$ show that 
when $\beta$ is large, configuration in 
Fig.\ \ref{fig5}(a) is the most favored 
and Fig.\ \ref{fig5}(c) is the least favored.
As $\beta$ is decreased, the favored configuration is changed
between $\beta$=0.5 and 0.7. 
After that Fig.\ \ref{fig5}(c) is the most favored and 
Fig.\ \ref{fig5}(a) is the least favored. 
When $\beta$ is large, 
the holes prefer to distribute evenly because 
of the same reason as in $x=1/8$ case. 
In contrast, when $\beta$ is small, 
electron sites prefer to have more neighboring electron sites to gain 
orbital energy.
Our results are not consistent with  
the experimental results 
for  ${\rm La_{1/2}Ca_{1/2}MnO_3}$,\cite{chen}
which indicate that configuration Fig.\ \ref{fig5}(b)
is the ground state.
This inconsistency 
may arise because our model involves only
localized electrons, 
while for $x=1/2$ the charge ordering state arises from a metallic phase.
Modifications of our model to include hole hopping
are desirable.  
 
In summary, we
have shown that the lattice effect 
could play an important role
in the charge ordering transition observed in perovskite
 manganites.

\section*{Acknowledgments}  

This work was supported in part by NSF-DMR-9705482.
We also acknowledge partial support from
NSF-DMR-96322526 (The Johns Hopkins M.R.S.E.C.
for Nanostructured materials).

\appendix
\section*{} 

To find the minimum energy 
we transform $E_{\rm tot}$ in Eq.\ (\ref{eq:TOT}) into $\vec{k}$ space,
using Eqs.\ (\ref{eq:ee}) $-$ (\ref{eq:orbital}). 
This leads to the following energy expressions in $k$ space:

\begin{eqnarray}
E_{\rm tot}/(NK_1)&=&
 \sum_{\vec{k}}
\delta_{\vec{k}}^{\dagger} M_{\vec{k}} \delta_{\vec{k}} +
\delta_{\vec{k}}^{\dagger} L_{\vec{k}}^{\dagger} u_{\vec{k}} +
u_{\vec{k}}^{\dagger} L_{\vec{k}} \delta_{\vec{k}} +
u_{\vec{k}}^{\dagger}  u_{\vec{k}} +
u_{\vec{k}}^{\dagger} P_{\vec{k}} e_{\vec{k}} +
e_{\vec{k}}^{\dagger} P_{\vec{k}}^{\dagger} u_{\vec{k}}  \nonumber  \\   
 & &- \frac{A}{NK_1} \sum_{i} (1-h_{i}) \cos{6\theta_i},  \label{eq:AEqE}
\end{eqnarray}
where 
 \begin{eqnarray}
\delta_{\vec{k}}^{\dagger} &=&  
( \delta^{x}_{\vec{k}}, \delta^{y}_{\vec{k}}, \delta^{z}_{\vec{k}} ), \\
u_{\vec{k}}^{\dagger} &=&
( u^{x}_{\vec{k}}, u^{y}_{\vec{k}}, u^{z}_{\vec{k}} ), \\
e_{\vec{k}}^{\dagger} &=&
( \beta h_{\vec{k}}-c^x_{\vec{k}} , \beta h_{\vec{k}}-c^y_{\vec{k}}, 
\beta h_{\vec{k}}-c^z_{\vec{k}} ), 
\end{eqnarray}

\begin{eqnarray}
&&M_{\vec{k}} =  
\left( 
   \begin{array}{ccc}
       \begin{array}{c} 
          1+\frac{K_2}{K_1}(1-\cos{k_x}) \\
          +\frac{K_3}{K_1}(1-\cos{k_x}\cos{k_y})  \\
            +\frac{K_3}{K_1}(1-\cos{k_x}\cos{k_z})
       \end{array} & 
       \frac{K_3}{K_1}\sin{k_x}\sin{k_y} &
       \frac{K_3}{K_1}\sin{k_x}\sin{k_z} \\ 
   \frac{K_3}{K_1}\sin{k_y}\sin{k_x} &
      \begin{array}{c}
      1+\frac{K_2}{K_1}(1-\cos{k_y}) \\
      +\frac{K_3}{K_1}(1-\cos{k_y}\cos{k_z}) \\
       +\frac{K_3}{K_1}(1-\cos{k_y}\cos{k_x}) \\
      \end{array} &
      \frac{K_3}{K_1}\sin{k_y}\sin{k_z} \\
   \frac{K_3}{K_1}\sin{k_z}\sin{k_x} &
      \frac{K_3}{K_1}\sin{k_z}\sin{k_y} &
      \begin{array}{c}
      1+\frac{K_2}{K_1}(1-\cos{k_z})  \\
      +\frac{K_3}{K_1}(1-\cos{k_z}\cos{k_x})   \\
      +\frac{K_3}{K_1}(1-\cos{k_z}\cos{k_y}) 
      \end{array}
   \end{array}
\right) ,
\end{eqnarray}

\begin{eqnarray}
L_{\vec{k}} &=& 
\left( \begin{array}{ccc}
   -\frac{1}{2}(1+e^{ik_{x}}) & 0 & 0 \\
    0 & -\frac{1}{2}(1+e^{ik_y}) & 0 \\
    0 & 0 & -\frac{1}{2}(1+e^{ik_z}) 
      \end{array}
\right), 
\\
P_{\vec{k}} &=&
\frac{\lambda}{2K_1}
\left( \begin{array}{ccc}
         1-e^{ik_{x}} & 0 & 0 \\
         0 & 1-e^{ik_y} & 0 \\
         0 & 0 & 1-e^{ik_z} 
       \end{array}
\right), 
\end{eqnarray}
and $N$ is the total number of Mn sites.
We obtain Eqs.\ (\ref{eq:nonzerok}) $-$ (\ref{eq:del}) 
by minimizing the above expression with respect to 
all $\delta^{a}_{\vec{k}}$ 
and 
$u^a_{\vec{k}}$. 
Without the second neighbor elastic energy term, 
$\delta_{\vec{k}}$ and $u_{\vec{k}}$ minimizing 
Eq.\ (\ref{eq:AEqE}) become singular
when any of $k_x$, $k_y$, and $k_z$ is zero.
With nonzero $K_3$ 
this singularity has been uniquely solved for 
$\vec{k}\neq 0$, 
while at $\vec{k}=0$, it is not. 

To find the energy term with $\vec{k}=0$,
we take the  $\vec{k}\rightarrow0$ limit.
That corresponds to the uniform strain energy, 
i.e., the energy related to the 
change of the lattice parameters from the original cubic structure.
Here the problem of the choice of the limiting process arises,
because the calculation shows 
that the different directions of the limiting process of 
$\vec{k}\rightarrow 0$ have given different
energies and different uniform strains.
Because the lower energy state is
favored after all, 
the appropriate limiting process will be the one 
which gives the minimum uniform strain energy, 
and it determines the uniform strain also.
When $K_3/K_1 \ll 1$, this appropriate limiting process has been
found to satisfy the condition of 
$k_x$, $k_y$, and $k_z\neq 0$.
As far as $k_x$, $k_y$, and $k_z$ are nonzero, 
the limits are different only 
in the order of $K_3/K_1$. 
Therefore, in the  $K_3/K_1\rightarrow 0$ limit,
any $\vec{k}\rightarrow0$ limit process satisfying the above condition
gives the correct expression of the minimum energy
term with $\vec{k}=0$.
It also gives a unique uniform strain.

\newpage

\begin{figure}
\caption{
The three hole ordering patterns for 
$Re_{7/8}Ak_{1/8}{\rm MnO_3}$ considered 
in our calculations.
Solid circles represent ${\rm Mn^{3+}}$, and open circles ${\rm Mn^{4+}}$.
}   
\label{fig1}
\end{figure}

\begin{figure}
\caption{ 
Spring constants: $K_1$ between the nearest neighbor Mn-O, 
$K_2$ between the first neighbor Mn-Mn,
and $K_3$ between the second neighbor Mn-Mn.  
}   
\label{fig2}
\end{figure}

\begin{figure}
\caption{
Phase diagram in $A/K_1$ versus $\beta$ plane for 
$Re_{7/8}Ak_{1/7}{\rm MnO_3}$.
$\lambda/K_1=0.045$, $K_2/K_1=0.5$, and $g=0$. 
}   
\label{fig3}
\end{figure}

\begin{figure}
\caption{
Phase diagram in $g$ versus $\beta$ plane 
for $Re_{7/8}Ak_{1/7}{\rm MnO_3}$.
$\lambda/K_1=0.045$, $K_2/K_1=0.5$, and $A/K_1=0.0002$.  
}   
\label{fig4}
\end{figure}

\begin{figure}
\caption{
The three hole ordering patterns for 
$Re_{1/2}Ak_{1/2}{\rm MnO_3}$ considered
in our calculations.
Solid circles represent ${\rm Mn^{3+}}$, and open circles ${\rm Mn^{4+}}$.
}   
\label{fig5}
\end{figure}

\newpage
\begin{table}
\caption{
Coordinates of site $i$, orbital states, ionic displacements,
and uniform strains
for the minimum energy configuration of Fig.\ \ref{fig1}(b), 
when 
$A/K_1=0.0002$, $\lambda/K_1=0.045$, $K_2/K_1=0.5$ and $\beta=2.5$.
}
\label{table1}
\begin{tabular}{c|c|c|c|c|c|c|c|c}
i & $(n_i^x,n_i^y,n_i^z)$ & $\theta_i({\rm radian})$ & 
    $\delta_i^x-\delta_{\vec{k}=0}^x$ & $\delta_i^y-\delta_{\vec{k}=0}^y$ &
    $\delta_i^z-\delta_{\vec{k}=0}^z$ &
    $u_i^x-u_{\vec{k}=0}^x$      & $u_i^y-u_{\vec{k}=0}^y$      
& $u_i^z-u_{\vec{k}=0}^z$         \\ \hline
1 &  (0,0,0) & hole site &  0  &  0  &  0  &  -0.135  
&  -0.134  &  -0.159  \\
2 &  (1,0,0)    &   1.11  &  0 & 0.007 & 0   & 0.135  & 0.007 & -0.039  \\
3 & (0,1,0)  & 1.97 &  0     & 0        &   0  &  0.004 
& 0.134 & -0.030  \\
4 & (1,1,0)  &   0.03 &     0 &        0 & 0  &  -0.004 
& -0.007 & 0.047  \\
5 & (0,0,1)  & 0.09 &  0  &  -0.005 & -0.049  &  0  
&  -0.023  &  0.011   \\
6 & (1,0,1) &  0.09 &   0  &  0.005 & -0.007  &  0  
&  -0.009 & 0.019  \\
7 & (0,1,1)  & 2.74 &  0  &  0  &  0.002  &  -0.043  &  0.023 & 0.037   \\
8 & (1,1,1) & 1.24 &   0   &   0   &  0.013  &  0.043  
&  0.009 & -0.008  \\
9 &   (0,2,0) & hole site &  0  &  0  &  0  &  -0.135  
& -0.133 & -0.159   \\
10 &  (1,2,0) & 1.11 &   0  &  -0.007 & 0   &  0.135  
& -0.013  & -0.039  \\
11 &  (0,3,0)   & 2.28 &  0  &  0  &  0  &  -0.036  &  0.133 &  0.008   \\
12 &  (1,3,0)  & 1.33 &  0  &  0  &  0  &  0.036 & 0.013  &  -0.037  \\
13 &  (0,2,1) & 0.09 &  0  &  0.005 &  -0.049  &  0  &  -0.009 &  0.011  \\
14 &   (1,2,1) &  0.09 &   0 &   -0.005 &  -0.007  &  0  
&  -0.023   &  0.019  \\
15 &  (0,3,1) & 1.24 &  0  &  0  &  -0.013   &  0.043  
&  0.009 & -0.047   \\
16 &  (1,3,1)  & 2.74 &  0 &  0  &  -0.002  & -0.043  
&  0.023  &  0.030  \\
\hline
\multicolumn{2}{c}{Uniform strain : }     &  \multicolumn{7}{c}
{ $e^{xx}=-0.014$,  $e^{yy}=-0.019$, $e^{zz}=-0.009$}    \\ 
\end{tabular}
\end{table}
\end{document}